\documentclass[preprint,5p]{elsarticle}

\usepackage{graphicx}

\usepackage{booktabs}
\usepackage{tabularx}
\usepackage{tikz}
\usepackage{pgfplots}
\usetikzlibrary{positioning,shapes,trees,arrows,graphs,decorations,calc}
\usepackage[absolute,overlay]{textpos}
\usepackage{url}

\usepackage{amssymb}
\usepackage{amsmath}
\usepackage{hyperref}
\usepackage{xcolor}

\usepackage{placeins} 
\usepackage[normalem]{ulem}

\biboptions{comma,compress}

\newcounter{bla}

\journal{Computer Physics Communications}

\begin{document}

\begin{frontmatter}



\title{{\tt MCNNTUNES}: tuning Shower Monte Carlo generators with machine learning}

\author[label1]{Marco Lazzarin}
\author[label2]{Simone Alioli}
\author[label1]{Stefano Carrazza}

\cortext[author] {Corresponding author.\\\textit{E-mail address:}
stefano.carrazza@unimi.it\\\textit{Preprint number:} TIF-UNIMI-2020-24}

\address[label1]{TIF Lab, Dipartimento di Fisica, Universit\`a degli Studi di Milano and
INFN Sezione di Milano, Milan, Italy.}
\address[label2]{Dipartimento di Fisica, Universit\`a degli Studi di Milano Bicocca and
INFN Sezione di Milano Bicocca, Milan, Italy.}

\begin{abstract}
The parameters tuning of event generators is a research topic characterized by
complex choices: the generator response to parameter variations is difficult to
obtain on a theoretical basis, and numerical methods are hardly tractable due to
the long computational times required by generators. Event generator tuning has
been tackled by parametrisation-based techniques, with the most successful one
being a polynomial parametrisation. In this work, an implementation of tuning
procedures based on artificial neural networks is proposed. The implementation
was tested with closure testing and experimental measurements from the ATLAS
experiment at the Large Hadron Collider.

\end{abstract}


\end{frontmatter}

\noindent
{\bf PROGRAM SUMMARY}
\\

\begin{small}
\noindent
{\em Program Title:} {\tt MCNNTUNES} \\
\\
{\em Program URL:} \url{https://github.com/N3PDF/mcnntunes}\\
\\
{\em Licensing provisions:} GPLv3 \\
\\
{\em Programming language:} {\tt Python} \\
\\
{\em Nature of the problem:} Shower Monte Carlo generators introduce many parameters that must be tuned to reproduce the experimental measurements.
The dependence of the generator output on these parameters is difficult to obtain on a theoretical basis.\\
\\
{\em Solution method:} Implementation of a tuning method using supervised machine learning algorithms based on neural networks, which are universal approximators.
 \\

\end{small}


\section{Introduction}
\label{sec:introduction}

Shower Monte Carlo (SMC) event generators are tools that simulate the collision
of particles at high energies. They introduce many parameters, mainly due to the
usage of phenomenological models, like the hadronization model or the underlying
event model, needed to describe the low-energy limit of QCD which is not easily
calculable from first principles. These parameters are difficult to obtain on a
theoretical basis, so they must be carefully tuned in order to make the
generators reproduce the experimental measurements. The procedure of estimating
the best value for each parameter is called event generator tuning.

This tuning procedure is made more difficult by the high computational cost of
running a generator, so it requires methods to study the dependence between a
generator output and its parameters. Moreover, since the observables considered
while analysing the generator output play a pivotal role in determining the
tuning, one needs to model this dependence for different observables at the same
time.

The current state-of-the-art tuning procedure is based on a polynomial
parametrisation of the generator response to parameter variations, followed by a
numerical fit of the parametrised behaviour to experimental data. This is the
procedure which is implemented in \texttt{Professor}~\cite{Buckley2009}, the
primary tool for SMC event generator tuning at the Large Hadron Collider (LHC).
However, the assumption that the dependence of the generator output on its
parameters is polynomial is not always justified.

This paper investigates new tuning procedures based on artificial neural
networks. Artificial neural networks are universal function approximators
\cite{Hornik1991,Leshno1993,UniversalApproximationTheoremWidth}, enabling fits
which are not biased towards polynomials, as shown in early
attempts~\cite{Andreassen:2019nnm}. Two different tuning procedures are
presented, called \emph{Per Bin} and \emph{Inverse} from now on. The former
follows the same approach of \texttt{Professor}, but with a different
parametrisation model made of fully-connected neural networks and a different
minimization algorithm: the evolutionary algorithm CMA-ES
\cite{CMA-ES_tutorial}. The latter takes a completely different approach: by
using a fully-connected neural network, it learns to predict directly the
parameters that the generator needs to output a given result. These two
procedures were implemented in the \texttt{Python} package
\texttt{MCNNTUNES}~\cite{stefano_carrazza_2020_4065208} and then tested with the
event generator \texttt{PYTHIA8} \cite{Sjostrand2015}. Two different datasets of
Monte Carlo runs were generated, with three and four tunable parameters
respectively. The procedures were tested with closure tests and with real
experimental data taken from the ATLAS experiment
\cite{ATLAS_ptZ_measurements,ATLAS_phi_measurements}.

A description of the procedures and their technical implementation is presented
in section \ref{sec:implementation}, while section \ref{sec:results} contains
the details of the testing phase. Finally, the conclusion of this work and
future development directions are presented in section \ref{sec:outlook}.

\section{Implementation}
\label{sec:implementation}
\texttt{MCNNTUNES} implements two different strategies for generator tuning, both based on feedforward neural networks. In this section these two strategies are presented in detail, along with a description of their technical implementation.

\subsection{\textit{Per Bin} model}
\label{subsec:per_bin_model}
The first strategy is a parametrisation-based method similar to \texttt{Professor} \cite{Buckley2009}. The work cycle is divided in three consecutive steps: dataset generation, parametrisation of the generator output, and the actual tuning step.

At first, a dataset of Monte Carlo runs is generated by sampling parameter configurations from the parameter space, and then running the generator with each configuration. This step is identical to the one in \texttt{Professor}.

\begin{figure}[tp]
    \centering
    \includegraphics[width=\linewidth]{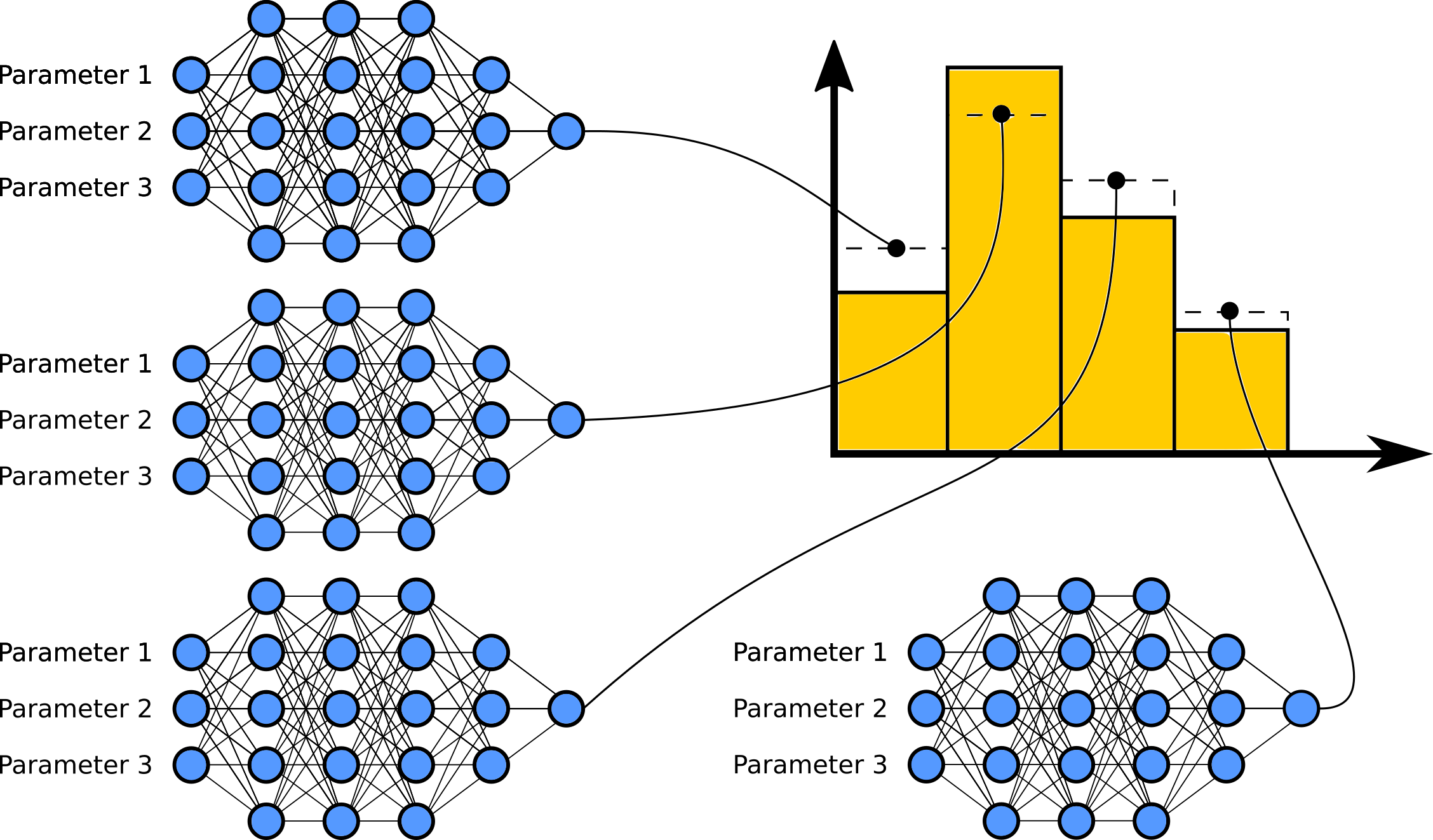}
    \caption{An illustration of the parametrisation of the generator response as implemented in the Per Bin Model.}
    \label{fig:per_bin_model}
\end{figure}

Then, the generator response to parameter variations is parametrised one bin at a time, using the previously created dataset.
The parametrisation is tackled by feedforward neural networks, which take the parameters as input and return the value of a single bin. An independent neural network is used for each bin\footnote{The hyperparameters are the same for each bin, only the parameters of the biases and the kernels are different.} (see Figure \ref{fig:per_bin_model} for an illustration).
The models are trained with a \mbox{gradient-based} algorithm, as usual for feedforward neural  networks, with mean squared error as loss. The details of the architecture, the choice of the optimization algorithm and its settings are all configurable by the user.

Finally, the tuning step exploits the parametric model of the generator to define a surrogate loss function for the tuning problem. In fact, the parametrisation step creates a model $h^{(i)}(\mathbf{p})$ of the generator, where $\mathbf{p}$ is the vector of parameters and $h^{(i)}$ the value of the $i$-th bin of the output. It enables the prediction of the generator output given a generic parameter configuration. The quality of the prediction will depend on the quality of the parametrisation. The optimization problem underlying the tuning can then be solved (approximately) by minimizing a surrogate loss function, e.g.
\begin{equation}
	\chi^2 (\mathbf{p}) = \sum_{i=1}^N \frac{\left( h^{(i)}(\mathbf{p}) - h_{\rm exp}^{(i)} \right)^2}{\sigma_{(i)}^2}
\end{equation}
The possibility of weighting each bin differently in the $\chi^2$ is also
implemented. Then, the minimization of this $\chi^2$ is performed with the
CMA-ES algorithm, which is a stochastic optimization method for non-linear
non-convex functions. The values of the parameters that minimize the $\chi^2$
determine the best tune. The minimization task can be carried out also with
gradient-based optimizers. This possibility is implemented as an optional
feature. A first test showed it performed worse than CMA-ES, but additional
tests with a careful tuning of the hyperparameters may change this result.

\subsection{Inverse model}
\label{subsec:inverse_model}
\begin{figure}[tp]
    \centering
    \includegraphics[width=\linewidth]{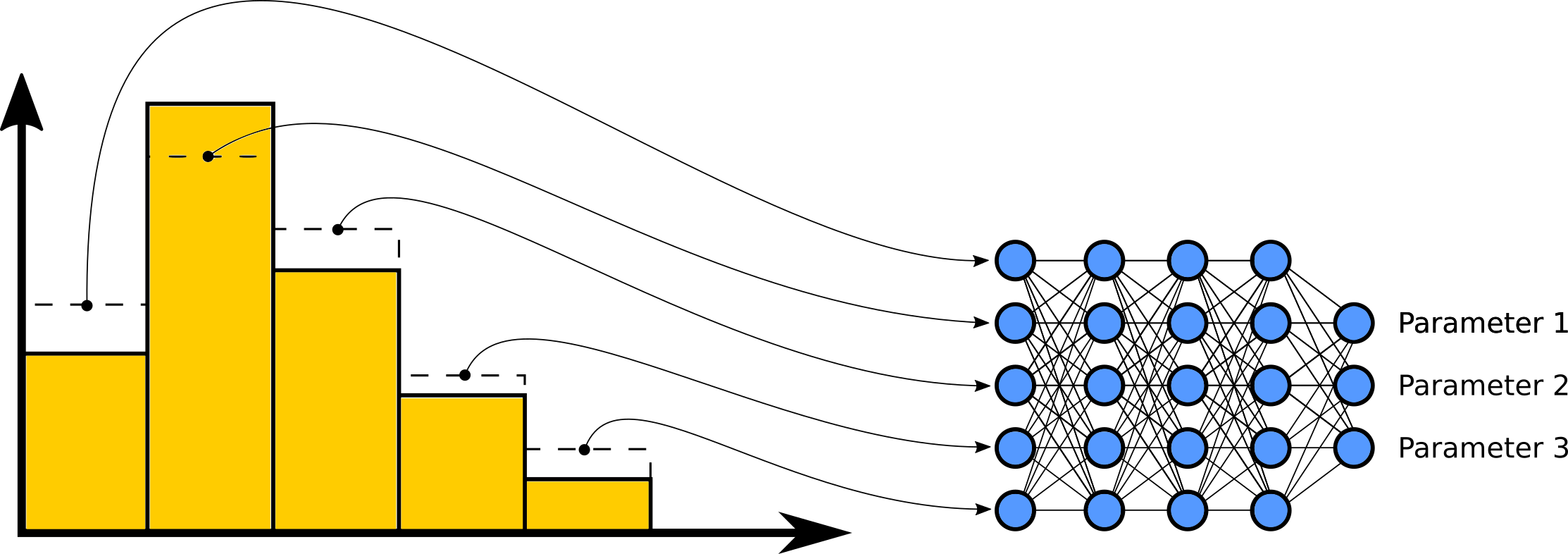}
    \caption{An illustration of the Inverse Model strategy.}
    \label{fig:inverse_model}
\end{figure}

The previous strategy involved models with parameters as input and histograms as output: given a set of parameters, it returns the histograms related to some observables. This is the input/output structure of a generator. In contrast, the Inverse model tries to learn the inverted model of a generator, using a feedforward neural network with the bin values as input layer and with the generator parameters as output layer (see Figure \ref{fig:inverse_model} for an illustration).
In case of success, the model is able to predict the parameters used for the generator given its results. Then, tuning the generator consists in feeding the experimental data into the model and inferring the parameters that the generator needs to reproduce them. The uncertainties of the predictions are computed in three steps:
\begin{enumerate}
    \item At first, the experimental data are resampled many times by using a multivariate Gaussian centered around the actual measurement, with a diagonal covariance matrix that includes the data uncertainties:
    $$\hbox{norm} \cdot \exp{ \left( - \frac{1}{2} \sum_{j=1}^{N_{\rm bins}} \left( (x_j - h_{j,\textrm{exp}})^2 / \sigma_{j,\textrm{exp}}^2 \right) \right) }$$
    \item This set of histograms is fed into the neural network.
    \item The output of the network is a distribution of predictions for each parameter, and the uncertainties are computed as the standard deviations of these distributions.
\end{enumerate}

\subsection{Data augmentation}
\label{subsec:data_augmentation}
During the training of the Inverse model the output variables (the parameters) are exact, but the input variables (the histogram bins) have a known uncertainty. In order to exploit this information, the training with jitter \cite{training_with_jitter} method was implemented as an optional feature: at each training epoch, the entire dataset is resampled following the data uncertainty. More precisely, let $\mathbf{X}$ be the dataset matrix where each row is a Monte Carlo run, and each column is the value of a bin, and let $\sigma_{ij}$ be the corresponding error for each element $X_{ij}$ of $\mathbf{X}$. Then, the training is done by replacing $\mathbf{X}$ with $\overline{\mathbf{X}}$ such that each element $\overline{X}_{ij}$ is a random variable distributed according to a Gaussian with mean $X_{ij}$ and variance $\sigma_{ij}^2$. $\mathbf{\overline{X}}$ is resampled at each epoch. This resembles a Gaussian noise layer applied to the input layer, but here the $\sigma$ of the Gaussian noise is different for each node of the input layer, and for each element of the training set. This method can be seen as a form of regularization, as proven in \cite{training_with_jitter}.

\subsection{Performance assessment}
\label{subsec:closure_tests}
The program implements a performance assessment procedure based on closure tests.
A single closure test consists in using one Monte Carlo run as the experimental data, and then performing the tune; in this way, the obtained tunes can be directly compared with the real parameters used to generate that run. Notice that that run must be excluded from the training set, otherwise the result does not measure the ability of the procedure to generalize to new examples.
The user can provide two different datasets of Monte Carlo runs: a training set, used to train the model, and a validation set, used to perform closure tests. Once the model has been trained, a closure test is performed for each run in the validation set, and a loss function defined as
\begin{equation}
\label{eq:closure_test_loss}
    L = \sum_i \frac{| p_i^{\rm true} - p_i^{\rm pred} |}{p_i^{\rm true}}
\end{equation}
is computed. The average of the losses of all closure tests is used as validation loss. This loss could be interpreted as an estimator of the accuracy of the tuning procedure, even though it is unsatisfactory: the experimental data and the Monte Carlo runs are not identically distributed nor generated by the same data-generating underlying process. In fact, the generator may be unable to represent the experimental data at all. However, this loss could be used to tune the hyperparameters of the model, or to compare different models.

\subsection{Hyperparameter tuning}
\label{subsec:implementation_hyperparameter_tuning}
\texttt{MCNNTUNES} features an hyperparameter search procedure implemented with \texttt{Hyperopt} \cite{Bergstra2013, Bergstra2015}. \texttt{Hyperopt} is a library dedicated to the hyperparameter optimization of machine learning algorithms. In particular, it takes care of scalar-valued functions whose arguments are defined over a search space with a potentially complicated structure: some arguments could be real-valued (e.g. learning rates), others could be discrete (e.g. the choice of the optimization algorithm), and the search space could be tree-structured, i.e. some variables are defined only when other parent variables take on a specific value (e.g. the number of hidden layers and the number of units of each hidden layer).

The implementation of \texttt{Hyperopt} requires the definition of a search
space and the definition of the function to minimize (the objective function).
The search space is provided by the user in the configuration of
\texttt{MCNNTUNES}, using the format specified in the documentation of
\texttt{Hyperopt}.

The objective function must receive the sampled hyperparameters as input, create
a model with these hyperparameters, train it, and evaluate the validation loss
of that model, or at least some sort of performance measurement with the ``lower
is better" format. \texttt{MCNNTUNES} computes the performance measurement
presented in section \ref{subsec:closure_tests} as validation loss, provided
that a valid validation set of Monte Carlo runs is available. Specifically, it
trains the model on the training set, then performs a closure test for every run
in the validation set, computes the loss in eq.~(\ref{eq:closure_test_loss}) for
each of them, and finally returns the average of these losses as validation
loss. \texttt{Hyperopt} implements two different algorithms: a random search,
and a Sequential Model-Based Optimization (SMBO) algorithm called Tree-structured
Parzen Estimator \cite{Bergstra2011}. \texttt{MCNNTUNES} uses the latter.

Moreover, \texttt{MCNNTUNES} supports the parallel search as implemented in \texttt{Hyperopt}.

\subsection{Technical details}
\label{subsec:technical_details}
The program is written in \texttt{Python}. The Monte Carlo runs (histograms) are loaded with the \texttt{YODA} library \cite{YODA}, which is the default histogram format of \texttt{Rivet} \cite{Rivet}. The basic operations are implemented with \texttt{NumPy} \cite{NumPy}, while the machine learning aspects use \texttt{Keras} \cite{Keras}, with the \texttt{TensorFlow} framework \cite{tensorflow2015-whitepaper} as backend. It uses the \texttt{pycma} package \cite{hansen2019pycma} for the CMA-ES algorithm.
The procedures are implemented in the \texttt{mcnntunes} script. The script accepts a configuration runcard in \texttt{YAML} format, which contains all program settings. This is the basic work cycle:
\begin{enumerate}
    \item \texttt{mcnntunes preprocess} loads the Monte Carlo runs and the experimental data, transforms the training set so that each input or output has mean 0 and variance 1, computes some useful statistics and saves all the data for future use.
    \item \texttt{mcnntunes model} trains the model specified in the runcard, and saves it for future use.
    \item \texttt{mcnntunes tune} performs the tune with the experimental data, and generates an \texttt{HTML} report with some information about the whole tuning process.
\end{enumerate}
Some additional features are useful for performance assessment (\texttt{mcnntunes benchmark}) and hyperparameter tuning (\texttt{mcnntunes optimize}).

\section{Results}
\label{sec:results}
This section presents the testing phase of \texttt{MCNNTUNES}. The choice of the
generator, the parameters with their variation ranges, the process and the
observables on which performing the tunes were chosen following the AZ tune
\cite{ATLAS_ptZ_measurements} as reference. This should be considered only as a
study of the efficiency and reliability of the \texttt{MCNNTUNES} approach for
some specific observables and data and not as an attempt to devise a new
exhaustive tune for the LHC.

The generation of some datasets of Monte Carlo runs is presented in section \ref{subsec:datasets}; a systematic performance assessment is presented in section \ref{subsec:validation_test_tuning}; finally, an AZ-like tune that tries to reproduce some results of \cite{ATLAS_ptZ_measurements} is presented in section \ref{subsec:final_tunes}.

\subsection{Datasets}
\label{subsec:datasets}
The generation of the datasets followed the procedure presented in \cite{ATLAS_ptZ_measurements}. The Monte Carlo runs were generated with \texttt{PYTHIA} version \texttt{8.240} \cite{Sjostrand2015}, interfaced with the \texttt{Rivet} \cite{Rivet} package, version \texttt{2.7.0}.
Two different analyses were performed: one involved the measurement of the $Z/\gamma^*$ boson transverse momentum distribution $p_T^Z$ in $pp$ collisions at $\sqrt{s} = 7$ TeV \cite{ATLAS_ptZ_measurements} (analysis \texttt{ATLAS\textunderscore 2014\textunderscore I1300647}), the other involved the measurement of angular correlation $\phi_\eta^\star$ \cite{ATLAS_phi_measurements} (analysis \texttt{ATLAS\textunderscore 2012\textunderscore I1204784}), which probes the same physics of $p_T^Z$ but with higher experimental resolution. Thus, the activated process was $f\overline{f} \to Z/\gamma^*$. The investigated parameters are the primordial $k_T$, the parton shower $\alpha_S(m_Z^2)$ and the parton shower damping factor for the lower $p_T$ cut-off (both for the initial state radiation, ISR from now on), and the damping factor for the lower $p_T$ cut-off for the multiparton interaction. Two different datasets were generated: one, the most similar to \cite{ATLAS_ptZ_measurements}, fixes the multiparton interaction parameter, while the other does not. The former will be the dataset 3P from now on, while the latter will be called 4P. The variation range of each parameter and the setup of \texttt{PYTHIA8} are presented in Table \ref{tab:setup_datasets}.
\begin{table}[htp]
    \centering
    \begin{tabular}{c c c}
    \toprule
    Parameter & Dataset 3P & Dataset 4P \\
    \midrule
    Primordial $k_T$ [GeV]  & 1.0 - 2.5         & 1.0 - 2.5 \\
    ISR $\alpha_S(m_Z^2)$   & 0.120 - 0.140     & 0.120 - 0.140 \\
    ISR $p_{T,0}^{\rm ref}$ [GeV] & 0.5 - 2.5         & 0.5 - 2.5 \\
    MPI $p_{T,0}^{\rm ref}$ [GeV] & 2.18 (fixed)      & 1.9 - 2.2 \\
    \midrule
    \texttt{PYTHIA8} base tune & tune 4C \cite{Tune4C} & tune 4C \cite{Tune4C} \\
    Number of events & $4 \cdot 10^6$ & $4 \cdot 10^6$ \\
    Number of runs & 512 & 1280 \\
    \bottomrule
    \end{tabular}
    \caption{\texttt{PYTHIA8} setup and variation ranges.}
    \label{tab:setup_datasets}
\end{table}

The tunes in this section rely in the reconstruction of the vector
boson properties by combining dressed leptons, as defined in the aforementioned
\texttt{Rivet} analyses. We limit ourselves to consider only distributions
inclusive in rapidity. The tunes are performed only for $p_T^Z < 26$ GeV and
$\phi_\eta^\star < 0.29$, unless ``all bins" is specified.
Three different sets of measurements are selected: one
with only $p_T^Z$ measurements, another with $\phi_{\eta}^\star$ measurements,
and one using only the muon channel $p_T^Z$ measurement and the electron channel
$\phi_\eta^\star$ measurement.

\subsection{Performance measurements}
\label{subsec:validation_test_tuning}
This subsection presents some performance measurements on \texttt{Professor} and \texttt{MCNNTUNES}. The procedure is described as follows.

At first, the dataset is splitted in training (80\%), validation (10\%) and test set (10\%).

Then, an hyperparameter optimization of the model is performed by training each hyperparameter configuration on the training set and selecting the configuration with the best loss computed on the validation set. The loss function is the one presented in subsection \ref{subsec:closure_tests}. For \texttt{Professor}, the hyperparameter search consisted in a grid search for the polynomial order. In practice, the best order was obtained by trying polynomials with degree from one to seven. The other options were kept at their default values, except the options \texttt{-s 2 --scan-n=100} for \texttt{prof2-tune}. For \texttt{MCNNTUNES}, the hyperparameter search was performed by running \texttt{Hyperopt}, feeding it with the validation loss. The \texttt{Hyperopt} configurations were the one in Table \ref{tab:hyperopt_conf_1_Inverse} plus another one focused on the architecture only, presented in Table \ref{tab:hyperopt_conf_4_Inverse}. Whether using data augmentation or not, instead, was chosen by performing a grid search on top of the \texttt{Hyperopt} scan.

Finally, the best model is retrained on both the training and the validation set, and its performance is evaluated by closure testing on the test set. A schematic view of the workflow is shown in Figure \ref{fig:workflow}.

The results are presented in Table \ref{tab:test_1_Inverse}, for the Inverse model only.
The performance of \texttt{MCNNTUNES} turns out to be slightly better than \texttt{Professor}, on average. Results are however limited to this particular benchmark, and may change with different random seeds, losses, datasets, parameters and observables. Moreover, this benchmark uses Monte Carlo runs, and not experimental data, so this performance measurement will not estimate the real tuning precision with real experimental data, because experimental data and Monte Carlo runs are not drawn from the same underlying data-generating distribution function.
Due to computation time constraints, performance measurements for the Per Bin model were limited to restricted observables and datasets, without validation-test split, but they showed solid results (see Table \ref{tab:hyperopt_search_1_PerBin}).

\begin{table}[htp]
    \centering
    \begin{tabular}{c c}
    \toprule
    Hyperparameter & Variation Range \\
    \midrule
    \# hidden layers & 2-5 \\
    Units per layer & 2-20 \\
    Activation function & tanh, relu, sigmoid \\
    Optimizer & various \texttt{Keras} optimizers \\
    Epochs & 250-10000 in discrete steps \\\
    Batch size & 100, 200, 300, 400, 500 \\
    \midrule
    Number of trials & 1000 \\
    \bottomrule
    \end{tabular}
    \caption{\texttt{Hyperopt} configuration for the Inverse Model - broad search.}
    \label{tab:hyperopt_conf_1_Inverse}
\end{table}
\begin{table}[htp]
    \centering
    \begin{tabular}{c c}
    \toprule
    Hyperparameter & Variation Range \\
    \midrule
    \# hidden layers & 3-4 \\
    Units per layer & 10-50 in step of 2 \\
    Activation function & sigmoid \\
    Optimizer & adam \cite{Adam} \\
    Optimizer learning rate & default value \\
    Initializer & Glorot uniform \cite{glorot} \\
    Epochs & 2500-15000 in steps of 500 \\
    Batch size & 128 \\
    \midrule
    Number of trials & 1000 \\
    \bottomrule
    \end{tabular}
    \caption{\texttt{Hyperopt} configuration for the Inverse Model - architecture only.}
    \label{tab:hyperopt_conf_4_Inverse}
\end{table}
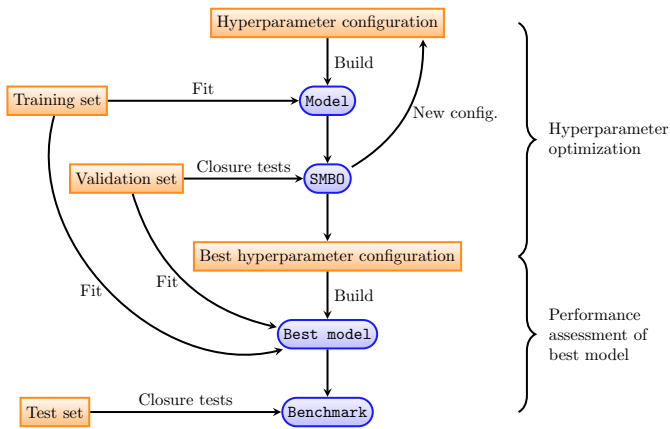
\begin{figure}[htp]
	\centering
	\resizebox{\columnwidth}{!}{
	\begin{tikzpicture}[
		nonterminal/.style={
				rectangle,
				minimum size=6mm,
				very thick,draw=orange!90, bottom color=orange!50,
				top color=orange!5},
		terminal/.style={
				rectangle,minimum size=6mm,
				rounded corners=3mm,very thick,
				draw=blue!90, bottom color=blue!25,
				top color=blue!5, font=\ttfamily},
				>=stealth,very thick
	]
		\node (hyperparameter_configuration)	[nonterminal]	{Hyperparameter configuration};
		\node (model) [terminal, below=of hyperparameter_configuration] {Model};
		\draw[->] (hyperparameter_configuration) edge node[midway,right] {Build} (model);
		\node (training)[nonterminal, left=of model, xshift=-3cm]	{Training set} ;
		\draw[->] (training) edge node[midway,above] {Fit} (model);

		\node (hp) [terminal, below=of model] {SMBO};
		\draw[->] (model) edge (hp);
		\node (validation)[nonterminal, left=of hp, xshift=-1.5cm] {Validation set};
		\draw[->] (validation) edge node[midway, above] {Closure tests} (hp);
		\coordinate(target) at ($(hyperparameter_configuration.center) + (2,-3.5mm)$);
		\draw[->, bend right=30] (hp) edge node[midway, right] {New config.} (target);

		\node (best_hp) [nonterminal, below=of hp] {Best hyperparameter configuration};
		\draw[->] (hp) edge (best_hp);
		\draw [decorate,decoration={brace,amplitude=10pt},xshift=3cm,yshift=1cm]
($(hyperparameter_configuration.center) + (4cm,0cm)$) -- ($(best_hp.center) + (4cm,0cm)$) node [black,midway,right,text width=2.5cm, align=left, xshift=0.5cm] {Hyperparameter optimization};

		\node (best_model) [terminal, below=of best_hp] {Best model};
		\node (test_error) [terminal, below=of best_model] {Benchmark};
		\node (test) [nonterminal, left=of test_error, xshift=-3cm] {Test set};
		\draw[->] (best_hp) edge node[midway,right] {Build} (best_model);
		\draw[->] (best_model) edge (test_error);
		\draw[->] (test) edge node[midway, above] {Closure tests} (test_error) ;
		\draw[->,bend right=30] (validation) edge node[midway,left] {Fit} (best_model);
		\draw[->,bend right=60] (training) edge node[midway,left] {Fit} (best_model);

		\draw [decorate,decoration={brace,amplitude=10pt},xshift=3cm,yshift=1cm]
($(best_hp.center) + (4cm,0cm)$) -- ($(test_error.center) + (4cm,0cm)$) node [black,midway,right,text width=2.5cm, align=left, xshift=0.5cm] {Performance assessment of best model};

	\end{tikzpicture}}
	\caption{Illustration of the performance assessment procedure.}
	\label{fig:workflow}
\end{figure}

\begin{table}[htp]
    \centering
    \begin{tabular}{l c c}
    \toprule
    Observables & Inverse (\%) & \texttt{Professor} (\%) \\
    \midrule
    $\phi_\eta^\star$, 3P                   & 3.7 $\pm$ 0.5 & 4.8 $\pm$ 0.7 (3)\\
    $p_T^Z$, 3P                             & 2.6 $\pm$ 0.3 & 3.1 $\pm$ 0.5 (3)\\
    $\phi_\eta^\star p_T^Z$, 3P             & 3.2 $\pm$ 0.4 & 3.6 $\pm$ 0.5 (3)\\
    $\phi_\eta^\star$, 3P, all bins         & 4.4 $\pm$ 0.6 & 4.2 $\pm$ 0.7 (3)\\
    $p_T^Z$, 3P, all bins                   & 2.5 $\pm$ 0.3 & 2.6 $\pm$ 0.4 (3)\\
    $\phi_\eta^\star p_T^Z$, 3P, all bins   & 3.1 $\pm$ 0.4 & 3.1 $\pm$ 0.5 (3)\\
    \midrule
    $\phi_\eta^\star$, 4P                   & 3.5 $\pm$ 0.2 & 4.6 $\pm$ 0.3 (5)\\
    $p_T^Z$, 4P                             & 2.71 $\pm$ 0.16 & 3.28 $\pm$ 0.18 (5)\\
    $\phi_\eta^\star p_T^Z$, 4P             & 3.2 $\pm$ 0.2 & 3.8 $\pm$ 0.3 (4)\\
    $\phi_\eta^\star$, 4P, all bins         & 3.7 $\pm$ 0.2 & 4.0 $\pm$ 0.3 (4)\\
    $p_T^Z$, 4P, all bins                   & 2.89 $\pm$ 0.15 & 3.2 $\pm$ 0.2 (4)\\
    $\phi_\eta^\star p_T^Z$, 4P, all bins   & 3.0 $\pm$ 0.2 & 3.4 $\pm$ 0.2 (4)\\
    \bottomrule
    \end{tabular}
    \caption{Test errors - Inverse Model against \texttt{Professor} (polynomial order inside parentheses).}
    \label{tab:test_1_Inverse}
\end{table}
\begin{table}[htp]
    \centering
    \begin{tabular}{c c c c}
    \toprule
    Observables & Per Bin (\%) & \texttt{Professor} (\%) \\
    \midrule
    $\phi_\eta^\star$, 3P & $3.1 \pm 0.5$ & $3.6 \pm 0.6$\\
    \midrule
    $\phi_\eta^\star$, 4P & $4.0 \pm 0.3$ & $4.0 \pm 0.3$\\
    \bottomrule
    \end{tabular}
    \caption{Validation losses (with smaller datasets). The hyperparameter search space was similar to the one of Table \ref{tab:hyperopt_conf_1_Inverse} but without configurations with five hidden layers.}
    \label{tab:hyperopt_search_1_PerBin}
\end{table}

\subsection{Tunes}
\label{subsec:final_tunes}
Finally, the datasets 3P and 4P were used to perform some final tunes. The hyperparameter configurations are the ones selected in the hyperparameter tuning step of subsection \ref{subsec:validation_test_tuning}, except for the Per Bin model for which the hyperparameters were chosen manually. The whole datasets were used for training. The results are presented in Table \ref{tab:3params_final_tune} and Table \ref{tab:4params_final_tune} for the 3P and 4P dataset respectively. Tunes obtained with the Per Bin model have no errors because no proper error estimation is implemented. A few comments on these tunes may be made:
\begin{itemize}
    \item \texttt{Professor} and the Per Bin model give similar results, usually compatible with each other.
    \item The Inverse model sometimes gives different results: the agreement with \texttt{Professor} is usually good for primordial $k_T$ and $\alpha_S^{ISR}(m_Z^2)$ parameters, with some exception. The other parameters are harder to analyse and will be discussed in the next points.
    \item The estimation of the MPI parameter by the Inverse model does not work: the model predicts always a parameter near the midpoint of the variation range, i.e. it is a trivial predictor. This did not prevent the Inverse model to perform better than \texttt{Professor} on closure testing, which means that both algorithms fail, and the trivial prediction is just the way the learning algorithm found to minimize the training loss.
    \item Unfortunately, many results suggest that the best value for ISR $p_{T,0}^{\rm ref}$ is somewhere outside the left bound of its variation range. This is easy to observe for two-steps methods like \texttt{Professor} and the Per Bin model: they model the generator behaviour in the parameter space, more precisely in the hyperrectangle populated by the dataset, while the tunes are found by a minimization algorithm that explores this hyperrectangle. When the tunes seem outside of the variation ranges the algorithm finds a minimum at the boundary of the parameters hyperrectangle. The minimizers can extrapolate the results outside of the variation ranges, but there the models may be unreliable. For the Inverse model it is more complicated, because the bounds are not hard-coded into the model. Moreover, it is difficult to understand if the experimental data are near some Monte Carlo runs, so that the prediction is reliable: the envelopes are not useful because they show only whether the experimental data are inside the bounding box of the Monte Carlo runs in histograms space, but the runs do not populate this bounding box uniformly. When \texttt{Professor} suggests a value outside the variation range, the behaviour of the Inverse model varies: sometimes it directly predicts a value outside the variation range, sometimes a value near the left bound, sometimes a value further away. When \texttt{Professor} founds a value inside the variation range, the corresponding value for the Inverse model is compatible with it (this happens in tunes performed over all bins, shown in Table \ref{tab:3params_final_tune_all_bins} and Table \ref{tab:4params_final_tune_all_bins}).
\end{itemize}
\begin{table*}
    \centering
    \begin{tabularx}{\textwidth}{l X X X}
    \toprule
    \multicolumn{4}{c}{\texttt{Professor}}\\
    Parameter               & $p_T^Z$               & $\phi_\eta^\star$     & $p_T^Z$ $\phi_\eta^\star$  \\
    \midrule
    Primordial $k_T$ [GeV]  & 1.77 $\pm$ 0.04       & 1.80 $\pm$ 0.04       & 1.77 $\pm$ 0.04\\
    ISR $\alpha_S(m_Z^2)$   & 0.1232 $\pm$ 0.0002   & 0.1236 $\pm$ 0.0002   & 0.1236 $\pm$ 0.0002\\
    ISR $p_{T,0}^{\rm ref}$ [GeV] & left bound            & left bound            & left bound\\
    \midrule
    \multicolumn{4}{c}{\texttt{MCNNTUNES}, Per Bin model}\\
    Parameter               & $p_T^Z$   & $\phi_\eta^\star$ & $p_T^Z$ $\phi_\eta^\star$  \\
    \midrule
    Primordial $k_T$ [GeV]  & 1.75          & 1.76          & 1.74 \\
    ISR $\alpha_S(m_Z^2)$   & 0.1233        & 0.1236        & 0.1236 \\
    ISR $p_{T,0}^{\rm ref}$ [GeV] & left bound    & left bound    & left bound \\
    \midrule
    \multicolumn{4}{c}{\texttt{MCNNTUNES}, Inverse model}\\
    Parameter               & $p_T^Z$               & $\phi_\eta^\star$     & $p_T^Z$ $\phi_\eta^\star$  \\
    \midrule
    Primordial $k_T$ [GeV]  & 1.75 $\pm$ 0.05       & 1.81 $\pm$ 0.05       & 1.77 $\pm$ 0.04 \\
    ISR $\alpha_S(m_Z^2)$   & 0.1249 $\pm$ 0.0006   & 0.1233 $\pm$ 0.0004   & 0.1241 $\pm$ 0.0005\\
    ISR $p_{T,0}^{\rm ref}$ [GeV] & 0.9 $\pm$ 0.2         & 0.24 $\pm$ 0.18       & 0.8 $\pm$ 0.2\\
    \bottomrule
    \end{tabularx}
    \caption{Tunes using dataset 3P.}
    \label{tab:3params_final_tune}
\end{table*}
\begin{table*}
    \centering
    \begin{tabularx}{\textwidth}{l X X X}
    \toprule
    \multicolumn{4}{c}{\texttt{Professor}}\\
    Parameter               & $p_T^Z$               & $\phi_\eta^\star$     & $p_T^Z$ $\phi_\eta^\star$  \\
    \midrule
    Primordial $k_T$ [GeV]  & 1.76 $\pm$ 0.05       & 1.80 $\pm$ 0.05       & 1.79 $\pm$ 0.04 \\
    ISR $\alpha_S(m_Z^2)$   & 0.1233 $\pm$ 0.0003   & 0.1237 $\pm$ 0.0002   & 0.1236 $\pm$ 0.0002 \\
    ISR $p_{T,0}^{\rm ref}$ [GeV] & left bound            & left bound            & 0.5 $\pm$ 1.9 \\
    MPI $p_{T,0}^{\rm ref}$ [GeV] & 2.11 $\pm$ 0.06       & 2.13 $\pm$ 0.07       & right bound \\
    \midrule
    \multicolumn{4}{c}{\texttt{MCNNTUNES}, Per Bin model}\\
    Parameter               & $p_T^Z$   & $\phi_\eta^\star$ & $p_T^Z$ $\phi_\eta^\star$  \\
    \midrule
    Primordial $k_T$ [GeV]  & 1.70      & 1.76              & 1.76 \\
    ISR $\alpha_S(m_Z^2)$   & 0.1233    & 0.1237            & 0.1236 \\
    ISR $p_{T,0}^{\rm ref}$ [GeV] & left bound & left bound       & left bound \\
    MPI $p_{T,0}^{\rm ref}$ [GeV] & 1.95      & right bound       & right bound \\
    \midrule
    \multicolumn{4}{c}{\texttt{MCNNTUNES}, Inverse model}\\
    Parameter               & $p_T^Z$               & $\phi_\eta^\star$     & $p_T^Z$ $\phi_\eta^\star$  \\
    \midrule
    Primordial $k_T$ [GeV]  &  1.69 $\pm$ 0.07      & 1.69 $\pm$ 0.04       & 1.60 $\pm$ 0.06 \\
    ISR $\alpha_S(m_Z^2)$   &  0.1246 $\pm$ 0.0007  & 0.12345 $\pm$ 0.00018 & 0.1238 $\pm$ 0.0007 \\
    ISR $p_{T,0}^{\rm ref}$ [GeV] & 0.9 $\pm$ 0.2         & 0.29 $\pm$ 0.09       & 0.6 $\pm$ 0.2 \\
    MPI $p_{T,0}^{\rm ref}$ [GeV] & 2.0468 $\pm$ 0.0011    & 2.0431 $\pm$ 0.0009    & 2.0450 $\pm$ 0.0009 \\
    \bottomrule
    \end{tabularx}
    \caption{Tunes using dataset 4P.}
    \label{tab:4params_final_tune}
\end{table*}

\begin{table*}
    \centering
    \begin{tabularx}{\textwidth}{l X X X}
    \toprule
    \multicolumn{4}{c}{\texttt{Professor}}\\
    Parameter               & $p_T^Z$               & $\phi_\eta^\star$     & $p_T^Z$ $\phi_\eta^\star$  \\
    \midrule
    Primordial $k_T$ [GeV]  & 1.82 $\pm$ 0.05       & 1.79 $\pm$ 0.04       & 1.74 $\pm$ 0.04\\
    ISR $\alpha_S(m_Z^2)$   & 0.1252 $\pm$ 0.0003   & 0.12370 $\pm$ 0.00017 & 0.1244 $\pm$ 0.0002\\
    ISR $p_{T,0}^{\rm ref}$ [GeV] & 1.27 $\pm$ 0.16       & left bound            & 0.80 $\pm$ 0.14\\
    \midrule
    \multicolumn{4}{c}{\texttt{MCNNTUNES}, Per Bin model}\\
    Parameter               & $p_T^Z$   & $\phi_\eta^\star$ & $p_T^Z$ $\phi_\eta^\star$  \\
    \midrule
    Primordial $k_T$ [GeV]  &  1.79     & 1.75              & 1.75 \\
    ISR $\alpha_S(m_Z^2)$   &  0.1251   & 0.1238            & 0.1246 \\
    ISR $p_{T,0}^{\rm ref}$ [GeV] &  1.18     & 0.54              & 0.89 \\
    \midrule
    \multicolumn{4}{c}{\texttt{MCNNTUNES}, Inverse model}\\
    Parameter               & $p_T^Z$               & $\phi_\eta^\star$     & $p_T^Z$ $\phi_\eta^\star$  \\
    \midrule
    Primordial $k_T$ [GeV]  & 1.87 $\pm$ 0.04       & 1.79 $\pm$ 0.03       & 1.75 $\pm$ 0.05 \\
    ISR $\alpha_S(m_Z^2)$   & 0.1256 $\pm$ 0.0003   & 0.12363 $\pm$ 0.00016 & 0.1244 $\pm$ 0.0003 \\
    ISR $p_{T,0}^{\rm ref}$ [GeV] & 1.36 $\pm$ 0.14       & 0.61 $\pm$ 0.08       & 0.85 $\pm$ 0.14 \\
    \bottomrule
    \end{tabularx}
    \caption{Tunes using dataset 3P, all bins.}
    \label{tab:3params_final_tune_all_bins}
\end{table*}
\begin{table*}
    \centering
    \begin{tabularx}{\textwidth}{l X X X}
    \toprule
    \multicolumn{4}{c}{\texttt{Professor}}\\
    Parameter               & $p_T^Z$               & $\phi_\eta^\star$     & $p_T^Z$ $\phi_\eta^\star$  \\
    \midrule
    Primordial $k_T$ [GeV]  & 1.80 $\pm$ 0.05       & 1.75 $\pm$ 0.04       & 1.69 $\pm$ 0.04 \\
    ISR $\alpha_S(m_Z^2)$   & 0.1253 $\pm$ 0.0003   & 0.12370 $\pm$ 0.00018 & 0.1241 $\pm$ 0.0003 \\
    ISR $p_{T,0}^{\rm ref}$ [GeV] & 1.33 $\pm$ 0.14       & left bound            & 0.5 $\pm$ 0.4 \\
    MPI $p_{T,0}^{\rm ref}$ [GeV] & 2.00 $\pm$ 0.06       & 2.01 $\pm$ 0.04       & 2.05 $\pm$ 0.04 \\
    \midrule
    \multicolumn{4}{c}{\texttt{MCNNTUNES}, Per Bin model}\\
    Parameter               & $p_T^Z$   & $\phi_\eta^\star$ & $p_T^Z$ $\phi_\eta^\star$  \\
    \midrule
    Primordial $k_T$ [GeV]  & 1.79      & 1.73              & 1.66 \\
    ISR $\alpha_S(m_Z^2)$   & 0.1252    & 0.1237            & 0.1241 \\
    ISR $p_{T,0}^{\rm ref}$ [GeV] & 1.32      & left bound        & left bound \\
    MPI $p_{T,0}^{\rm ref}$ [GeV] & 1.90      & 1.99              & 2.01 \\
    \midrule
    \multicolumn{4}{c}{\texttt{MCNNTUNES}, Inverse model}\\
    Parameter               & $p_T^Z$               & $\phi_\eta^\star$     & $p_T^Z$ $\phi_\eta^\star$  \\
    \midrule
    Primordial $k_T$ [GeV]  & 1.79 $\pm$ 0.05       & 1.90 $\pm$ 0.06       & 1.72 $\pm$ 0.04 \\
    ISR $\alpha_S(m_Z^2)$   & 0.1250 $\pm$ 0.0003   & 0.1231 $\pm$ 0.0002   & 0.1238 $\pm$ 0.0003 \\
    ISR $p_{T,0}^{\rm ref}$ [GeV] & 1.12 $\pm$ 0.15       & 0.526 $\pm$ 0.016     & 0.83 $\pm$ 0.13 \\
    MPI $p_{T,0}^{\rm ref}$ [GeV] & 2.0473 $\pm$ 0.0005   & 2.04411 $\pm$ 0.00016 & 2.0460 $\pm$ 0.0004 \\
    \bottomrule
    \end{tabularx}
    \caption{Tunes using dataset 4P, all bins.}
    \label{tab:4params_final_tune_all_bins}
\end{table*}

\section{Outlook}
\label{sec:outlook}
A deep learning approach to event generator tuning was presented by introducing
two different procedures, called Per Bin strategy and Inverse strategy
respectively. The former is a variation of the \texttt{Professor} tuning
procedure, that improves over it by relaxing the assumption of a polynomial
dependence of the generator response to variations of the parameters. The latter
is a novel and completely different approach. The procedures were tested with
closure tests and real experimental data, though in low dimensional parameter
spaces. The Per Bin model closure testing was very limited, due to computational
time constraints, but showed solid results. The test with real experimental data
showed a behaviour similar to the one of \texttt{Professor}. The Inverse model
closure testing presented slightly better performances than the ones of
\texttt{Professor}, while the test with real experimental data showed some
differences from the other procedures.

In addition to the fact that the parametrisation is not bound to polynomials
anymore, already mentioned, another advantage of \texttt{MCNNTUNES} is that the
models can learn highly non-linear functions, at least in principle. This can be
of support to the Inverse model strategy: a two-step method of parametrisation
and minimization is replaced by a single-step one, which is conceptually
simpler, but the function to learn is more complicated.

On the other hand, the procedure brings all the difficulties that are typical of
deep learning algorithms: the complexity of the training step, the dependence of
the performance on the choice of the hyperparameters, the difficulty in the
interpretation of the behaviour of the trained model, the overfitting problem.
Moreover, the hyperparameter tuning is computationally expensive, especially for
the Per Bin model, and this prevents the models to reach their full potential.
Finally, the Inverse strategy introduces some practical problems, e.g. the error
estimation and the reliability of the predictions when the experimental
measurements have no Monte Carlo runs near them.

The behaviour of the procedures with a wider set of experimental data is still unclear,
and requires more in-depth studies. In addition, whether the procedures scale
well with the number of parameters is still to be determined. Future
investigations may involve studying their performances in high dimensional
parameter spaces. Finally, further developments may solve the practical problems
highlighted above.

Nevertheless, this is a first attempt to bridge the power of machine learning
algorithms into the complexity of SMC tuning. We auspicate that the greater
flexibility allowed by this tool will facilitate the tuning efforts inside the
experimental collaborations.

\section*{Acknowledgements}

This research used resources of the National Energy Research Scientific
Computing Center (NERSC), a U.S. Department of Energy Office of Science User
Facility operated under Contract No. DE-AC02-05CH11231. S.C. is supported by the
European Research Council under the European Union's Horizon 2020 research and
innovation Programme (grant agreement number 740006). The work of S.A. is
supported by the ERC Starting Grant REINVENT-714788. He also acknowledges
funding from Fondazione Cariplo and Regione Lombardia, grant 2017-2070 and by
the Italian MIUR through the FARE grant R18ZRBEAFC.

\bibliographystyle{elsarticle-num}
\bibliography{mcnntunes.bib}

\end{document}